\documentclass{PoS}

\title{The RooStats Project}

\ShortTitle{RooStats}

\author{\speaker{L. Moneta } \\
 CERN, Geneva, Switzerland \\
 E-mail: \email{Lorenzo.Moneta@cern.ch} }

 \author{K. Belasco \\
        Princeton University, USA
      }       
 \author{K. S. Cranmer \\
       New York University, USA
      }       
 \author{S. Kreiss \\
       New York University, USA
      }       
\author{A. Lazzaro \thanks{Speaker replacing L. Moneta} \\
 CERN, Geneva, Switzerland 
}    
  \author{D. Piparo \\
  KIT - Karlsruhe Institute of Technology, Germany
 }
\author{G. Schott \\
        KIT -  Karlsruhe Institute of Technology, Germany
     }     
 \author{W. Verkerke\\
       NIKHEF - National Institute for Subatomic Physics, Netherlands
       }
  \author{M. Wolf\\
    KIT - Karlsruhe Institute of Technology, Germany
}

\abstract{
RooStats is a project to create advanced statistical tools required for the analysis of 
LHC data, with emphasis on discoveries, confidence intervals, and combined measurements.
The idea is to provide the major statistical techniques as a set of C++ classes with coherent interfaces, so that can be used on arbitrary model and datasets in a common way. The classes are built on top of the RooFit package, which provides functionality for easily creating probability models, for analysis combinations and for digital publications of the results. 
We will present in detail the design and the implementation of the different statistical methods of RooStats.  We will describe the various classes for interval estimation and for hypothesis test depending on different statistical techniques such as those based on the likelihood function, or on frequentists or bayesian statistics. These methods can be applied in complex problems, including cases with multiple parameters of interest and various nuisance parameters. 
          }

\FullConference{13th International Workshop on Advanced Computing and Analysis Techniques in Physics Research, ACAT2010\\
		February 22-27, 2010\\
		Jaipur, India}

\begin{document}

\section{Introduction}

The Large Hadron Collider (LHC), which started to take data at the end of 2009, is expected to produce an unprecedented amount of data to be analyzed. 
These data offer a tremendous potential for the discovery of new physics and pose many challenges to the statistical techniques used in High Energy Physics (HEP). 
Different statistical approaches will be used for statistical inference of the data. 
That implies different techniques to be considered, each one answering different questions and using specific treatment of systematic uncertainties. 
Furthermore, statistical combinations of analyses of different search channels from a single  experiment  and combination of results from different experiments will be required. 
Up to now, in past experiments like those at LEP or at the Tevatron, dedicated code specific to the statistical method and to the analysis has been used to produce a result (for example an exclusion limit)  and to perform the combinations.

For the LHC experiments,  there has been a need to develop a generic and versatile software for implementing the required statistical tools. The code should be able to use, for the same input model, different statistical methods. The software needs also to be versatile in order to be able to cope with both simple analyses, such as those based on number counting, and complex ones which use the parametrization of experimental distributions. 

Three major types of statistical techniques have been identified: 
\begin{itemize}
 \item {\bf Classical / Frequentist}: this ``school'' of statistics restricts itself to making statements of the form ``probability of the data given the hypothesis''.  The definition of probability in this context is based on a limit of frequencies of various outcomes.  In that sense it is objective.
 \item {\bf Bayesian}: this ``school'' of statistics allows one to make statements of the form "probability of the hypothesis given the data", which requires a prior probability of the hypothesis.  Often the definition of probability in this context is a ``degree of belief''.  
 \item {\bf Likelihood-based}: this intermediate approach also uses a frequentist notion of probability (i.e. does not require a prior for the hypothesis), but it does not guarantee  to have the properties (i.e. coverage) that frequentists methods aim to achieve (or achieve by construction). This approach does ``obey the likelihood principle'' (as do Bayesian methods), while frequentist methods do not. 
\end{itemize}

These methods differ also in the way they incorporate the nuisance parameters for producing the results, so there is a need to have all of them. 

In order to fulfill the previously described requirements, a new project, RooStats, was started at the end of 2008, 
merging previous code developed by Kyle Cranmer for Phystat 2008~\cite{kyle} and from a CMS project, RooStatsCms~\cite{roostats_cms}.
RooStats is organized as a joint collaboration between ATLAS and CMS and is based on ROOT and RooFit. The developments are overseen by the ATLAS and CMS statistical committees. 
It has been built on top of ROOT~\cite{ROOT}, since ROOT is the most widely used tool in HEP for data analysis and provides the needed basic software (mathematical libraries, graphics libraries for plotting, I/O and dictionary libraries). RooStats is also based on the classes of the RooFit toolkit ~\cite{roofit}, which allows a convenient description of the data model and is a well-established software within the HEP community. 

All statistical methods require a probability density function and/or a likelihood function as input model. 
RooFit provides classes for facilitating the declaration of models and a way to re-use them in multiple statistical methods, since it has no statistical notion of parameters and observables. Therefore, it can work naturally with both frequentist and bayesian techniques.
RooStats can then be considered as providing high-level statistical tools, while RooFit provides the core data modeling infrastructure.

\subsection{Type of Statistical Questions}
One of the first steps in any statistical analysis is to carefully pose the question that one wishes to answer.  Most of these questions can be classified as follows:
\begin{itemize}
 \item {\bf Parameter Estimation}: find the most likely (`best fit') values of the parameters of a model, given the data.
 \item {\bf Hypothesis Testing}:  based on the data accept or reject a given hypothesis.  Often one tests a null hypothesis against an alternative.  When the hypothesis has no free parameters it is called `simple' and when it has free parameters it is called `composite'.
 \item {\bf Confidence intervals}: find a region in the parameter space that is consistent with the data.  In the frequentist setting, one desires for this interval or region to `cover' the true parameter point with a specified probability (or confidence).  In the Bayesian setting, one wishes for the interval or region to contain some fixed amount of the posterior probability.
 \item {\bf Goodness of Fit}: quantify how well a model fits the data, without reference to an alternative.
\end{itemize}
RooFit provides the functionality for parameter estimation while
RooStats provides tools for confidence interval estimations and hypothesis testing.  Classes for goodness of fit are currently provided in the core ROOT mathematical libraries\footnote{see the ROOT functions {\tt TH1::Chi2Test} and {\tt TMath::KolmogorovTest}}.

\section{RooStats Interfaces}

The fundamental design philosophy adopted for RooStats reflects a clear mapping between 
mathematical concepts and software implementation. 
We have tried to identify the fundamental concepts in statistics and then we assign them the corresponding C++ classes or interfaces.
The design of the interfaces follows the schema given by the statistical questions shown above.

\begin{figure}[htbp]
\begin{center}
\includegraphics[width=\textwidth]{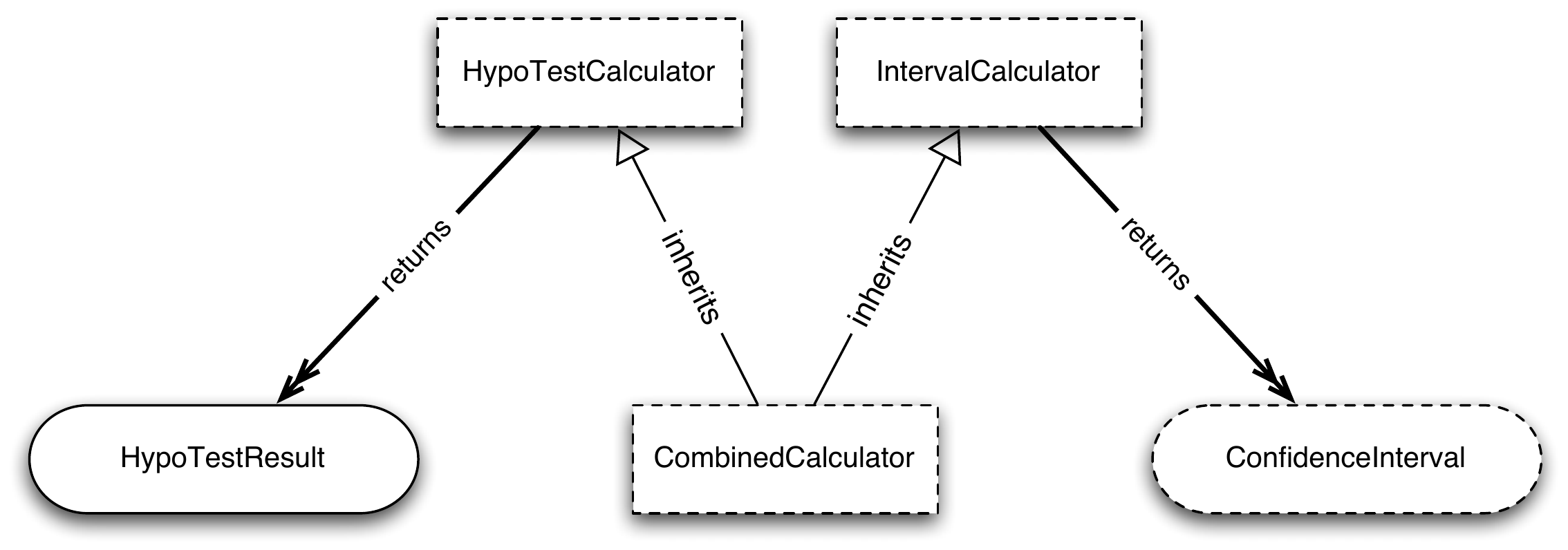}
\caption{A class diagram of the interfaces for hypothesis testing and confidence interval calculations.  The diagram shows the classes used to return the results of these statistical tests as well.}
\label{fig:OverviewOfInterfaces}
\end{center}
\end{figure}

\par
As shown in figure~\ref{fig:OverviewOfInterfaces},  RooStats provides a common interface for confidence interval calculations,  {\tt IntervalCalculator}, and another interface for performing the hypothesis tests,  {\tt HypoTestCalculator}. 
\par
The  {\tt IntervalCalculator} produces, as result, a confidence interval, which is represented by the RooStats {\tt ConfInterval} interface. 
The resulting type of {\tt ConfInterval} will depend on the implementation used for the interval calculator. There are many types of intervals, they can be simple one dimensional intervals or disconnected  regions in multiple dimensions. The common interface provides just the functionality to query whether a given point is inside 
the interval.
The {\tt IntervalCalculator} interface allows the user to set the  input model, the data set, the parameters of interest, the nuisance parameters and the confidence level or size of the test. 
\par
The  {\tt HypoTestCalculator} is the interface for performing hypothesis tests. The user gives as input  the model, the data set and the parameter lists specifying the null and the alternate hypothesis. The result of the class is 
an object implementing the {\tt HypoTestResult} interface. From this interface the user can retrieve the p values of the null and alternate hypothesis, the confidence level values and the corresponding significance, defined in terms of one-sided Gaussian standard deviations.

\section{RooStats Calculators}

The interfaces described above are then implemented in RooStats by different classes depending on the statistical techniques used. We now describe briefly all these statistical methods and their corresponding implementations as RooStats classes.
For a full description of the RooStats classes, see the 
RooStats reference documentation~\footnote{see http://root.cern.ch/root/htmldoc/ROOFIT\_ROOSTATS\_Index.html }.

\subsection{ProfileLikelihood Calculator}

\begin{figure}[hbt]
  \begin{center}
    \includegraphics[width=0.9\textwidth]{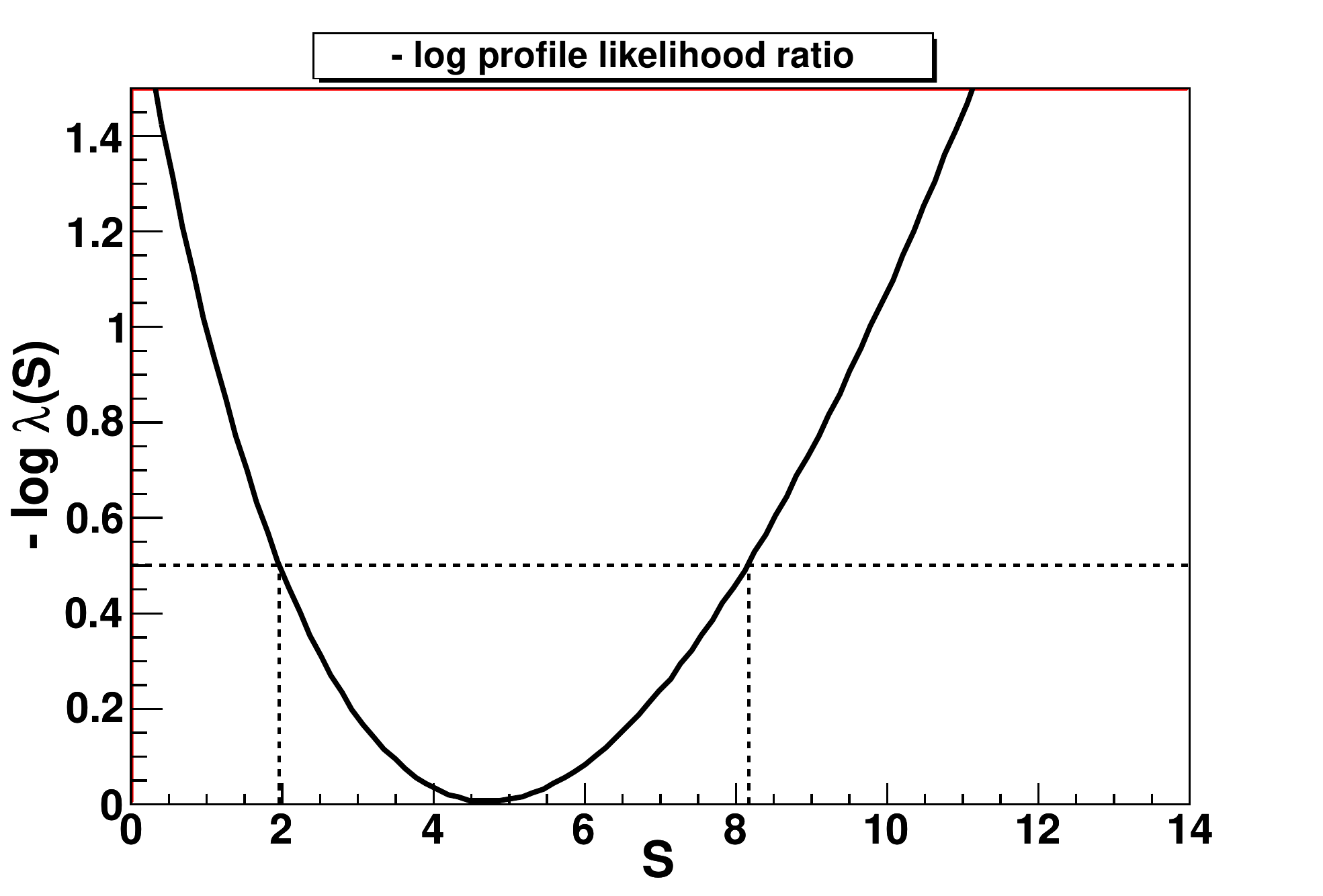} 
        \caption{
        Plot of the log profile likelihood curve ($-\log \lambda$ ) as function of the parameter of interest, S. The one $\sigma$ interval ($66.8\%$ CL) is obtained from  the intersect of the $-\log \lambda$  curve with the horizontal dashed line  ( $-\log \lambda = 0.5$).  
                 \label{likelihood_plot} }
  \end{center}
\end{figure}

This  calculator implements a likelihood-based method to estimate a confidence level and to perform an hypothesis test for a given parameter value.  
The likelihood function can be defined by the equation:
\begin{equation}
    \label{likelihood}
    L(\underline{x},\underline{\theta})=\prod^{N}_{i=1}f({\underline{x}}_{i},\underline{\theta}),
\end{equation}
where   \mbox{$\underline{x}=(x_a,\,x_b,\,x_c,\,\ldots)$} is a set of $N$ 
measured quantities, whose distributions are described by a joint probability density
function, $f(\underline{x},\underline{\theta})$, where
$\underline{\theta}=(\theta_1,\,\theta_2,\,\theta_3,\,\ldots )$ is a
set of $K$ parameters. 
For simplicity, let's focus on one dimensional case where  we are have
a parameter of interest,  $\theta_{0}$ and $K-1$ nuisance parameters. 
The profile likelihood is a function of $\theta_{0}$ and it is defined as 
\begin{equation}
    \label{profile_likelihood}
\lambda(\theta_{0} ) = \frac{
  L(\theta_{0},\hat{\hat{\underline{\theta} } }_{i \neq 0}
     ) } 
{  L(\hat{\theta}_{0}, \hat{\underline{\theta}}_{i \neq
      0}   ) }
\end{equation}
The denominator, $ L(\hat{\theta}_{0} )$ is the maximum likelihood value,
obtained from the best fit to the data, while the numerator is the
maximum value obtained by fixing $\theta_{0}$ and varying the remaining
$K-1$ parameters.
It can be shown (Wilks's theorem) that asymptotically
$-2\ln\lambda(\theta_0)$ is distributed as a $\chi^2(1)$ distribution ($\chi^2(n)$ 
in the case of $n$ parameter of interest). This is not
surprising since in the asymptotic limit the Likelihood function
becomes a Gaussian centered around the maximum likelihood estimator $\hat{\theta}_{0}$, and
the profile likelihood curve has a parabolic shape:
\begin{equation}
    \label{parabolicnll}
-2\ln\lambda(\theta_0) = -2 (\ln{L(\theta_0)}-\ln{L(\hat{\theta}_0)}) = n_{\sigma}^{2},\ \mathrm{with}\ n_\sigma=\frac{\theta_{0}-\hat{\theta}_{0}}{\sigma}. 
\end{equation}
where $\sigma$ represent the Gaussian standard deviations of the
parameter $\theta_{0}$. From this construction, it is possible to
obtained the one- or two-sided confidence interval we are
interested in (see for example figure~\ref{likelihood_plot}).  Even in case of non parabolic log-likelihood
functions, it can be shown, due to the invariance property of the
likelihood ratios, that this approach is still valid. This method is also called MINOS in the physics community, 
since it is implemented by the MINOS algorithm of the Minuit
program~\cite{Minuit}. 

Given the fact that asymptotically $-2\ln\lambda$ is distributed as  a $\chi^2$ distribution, an hypothesis tests can also be performed for 
the null hypothesis $H_0$, corresponding to $\theta = \theta^{\prime}$, and the alternate hypothesis $H_1$ for $\theta \neq \theta^{\prime}$.  
\par
This likelihood-based technique for estimating an interval and performing an hypothesis test is provided in RooStats  by the  {\tt ProfileLikelihoodCalculator}  class.  The class implements both the 
{\tt IntervalCalculator} and the {\tt HypoTestCalculator} interfaces.  When estimating the interval, it  
returns as result  a {\tt LikelihoodInterval} object, which, in the case of multi-parameters of interest, represents 
a multi dimensional contour. When performing the hypothesis test, an {\tt HypoTestResult} is returned with the 
significance and p-value for the null hypothesis. 
 Another class exists, {\tt LikelihoodIntervalPlot}, to visualize the likelihood interval in the case of one or two dimension (figure~\ref{likelihood_plot} shows the obtained plot in one dimensional problem).

\subsection{Bayesian Calculators}

These calculators are based on the Bayes theorem to estimate the posterior probability distribution for the desired parameters of interest. From the posterior probability distribution the  Bayesian credible interval can then be obtained. Input to the calculators are the model and the data sets which are used to build the likelihood function and  also the prior distributions of the parameters of interest and of the nuisance parameters, if those are present. 
These are integrated (marginalized) by each Bayesian calculator to obtained the posterior distribution.
RooStats provide two different types of Bayesian calculators, the {\tt BayesianCalculator} class and {\tt MCMCCalculator}, depending on the method used for performing the integration. 
 
The {\tt BayesianCalculator} class works only for one single parameter of interest and it uses analytical or numerical integration to compute the posterior probability. The result of the class is a one dimensional interval (class {\tt SimpleInterval} ), which is obtained from the cumulative posterior distribution.  

The {\tt MCMCCalculator} uses a Markov-Chain Monte Carlo to perform the integration. It runs the Metropolis-Hastings algorithm to construct the Markov Chain.  The class allows to configure the Metropolis-Hastings algorithm, via number of interations and burn-in-steps. Moreover, it is possible to replace the default uniform proposal function with any other distribution function. The result of the  {\tt MCMCCalculator} is a {\tt MCMCInterval}, which can compute the confidence interval on a desired parameter of interest from the obtained Markov Chain.  {\tt MCMCInterval} integrates the posterior where it is the tallest until it finds the correct cut-off height $C$ to give the target confidence level $1-\alpha$, such that
	\begin{center}
	$\displaystyle \int\limits_{f(\mathbf{x}) \, \ge \, C} f(\mathbf{x}) \,d^nx = 1-\alpha$.
	\end{center}

To perform this calculation, the posterior can be represented using  an histogram in one or multi-dimensions or using a kernel estimator.  The class  {\tt MCMCIntervalPlot} class can be used to visualize the interval and the Markov chain. 

Users can input the RooStats model also to the Bayesian Analysis Toolkit (BAT) ~\cite{bat}, a different software package implementing Bayesian methods with the Markov-Chain Monte Carlo. In its latest release, BAT provides a class, {\tt BATCalculator} which can be used with a similar interface to the RooStats  {\tt MCMCCalculator} .

\subsection{Neyman Construction}

The Neyman construction is a pure frequentist method to construct an interval at a given confidence level value, $1-\alpha$, in order that  if we repeat the experiment many times the interval will contain the true value a fraction $1-\alpha$ of the time. A detailed description of the method is given in~\cite{James_book}. 
RooStats provides a class, {\tt NeymanConstruction} implementing this technique. The class derives from  {\tt IntervalCalculator} and  returns as result  a {\tt PointSetInterval}, 
a concrete implementation of {\tt ConfInterval}. 

The Neyman Construction is not a uniquely defined statistical technique, it requires that one specifies an ordering rule or ordering principle, which is usually encoded by choosing a specific test statistic and limits of integration (corresponding to upper/lower/central limits). As a result, the RooStats class must be configured with the corresponding information before it can produce an interval. 
These have been generalized with the interfaces {\tt TestStatistic}, {\tt TestStatSampler}, and {\tt SamplingDistribution}.
Common configurations, such as the unified Feldman-Cousins approach, where the ordering is based on the profile likelihood ratio as test statistics and it is described in their paper~\cite{fc_paper},  can be enforced by using the {\tt FeldmanCousins} class. 

The Neyman construction considers every point in the parameter space independently, no assumptions are made that the interval is connected or of a particular shape. The result consists then of a set of scanned points with the information if they are inside or outside the interval ( {\tt PointSetInterval} class). The user indicates which points in the parameter space to perform the construction or provides a range with  the number of points which will be scanned  uniformly in a grid. 
For each scanned point,  the calculator will obtain the sampling distribution of the chosen test statistics. This is typically obtained by toy Monte Carlo sampling, but other techniques exist and they can in principle be used, although there are not yet implemented in RooStats.   
The tool also implements a generalization of the Feldman-Cousins procedure when nuisance parameters are present as described in~\cite{kyle,Cranmer2003}.

\subsection{Hybrid Calculator}

This calculator implements an 
hybrid approach for hypothesis testing. It consists of a traditional frequentist toy Monte Carlo method, 
like in the Neyman construction, but  with a 
Bayesian marginalization of nuisance parameters~\cite{cousins-highland}.
Hence the technique is often referred to as a
``Bayesian-Frequentist Hybrid''.
For example,  we define the null hypothesis, $H_b$, that no signal is
present over the background and $H_{sb}$ the alternate hypothesis that signal is
also present. In order to quantify the degree to which each hypothesis
is favored or excluded by the experimental observation, one chooses a
test-statistics which ranks the possible experimental outcomes. A
commonly used test statistics is the ratio of the likelihood
function in both hypotheses: $Q=L_{sb}/L_b$ and the quantity $-2\ln Q$
may also be used instead of $Q$. 
Alternative choices of test statistics can be the number of events or 
the profiled likelihood ratio (see equation~\ref{profile_likelihood}).

\begin{figure}[htb]
  \begin{center}
    \includegraphics[width=0.9\textwidth]{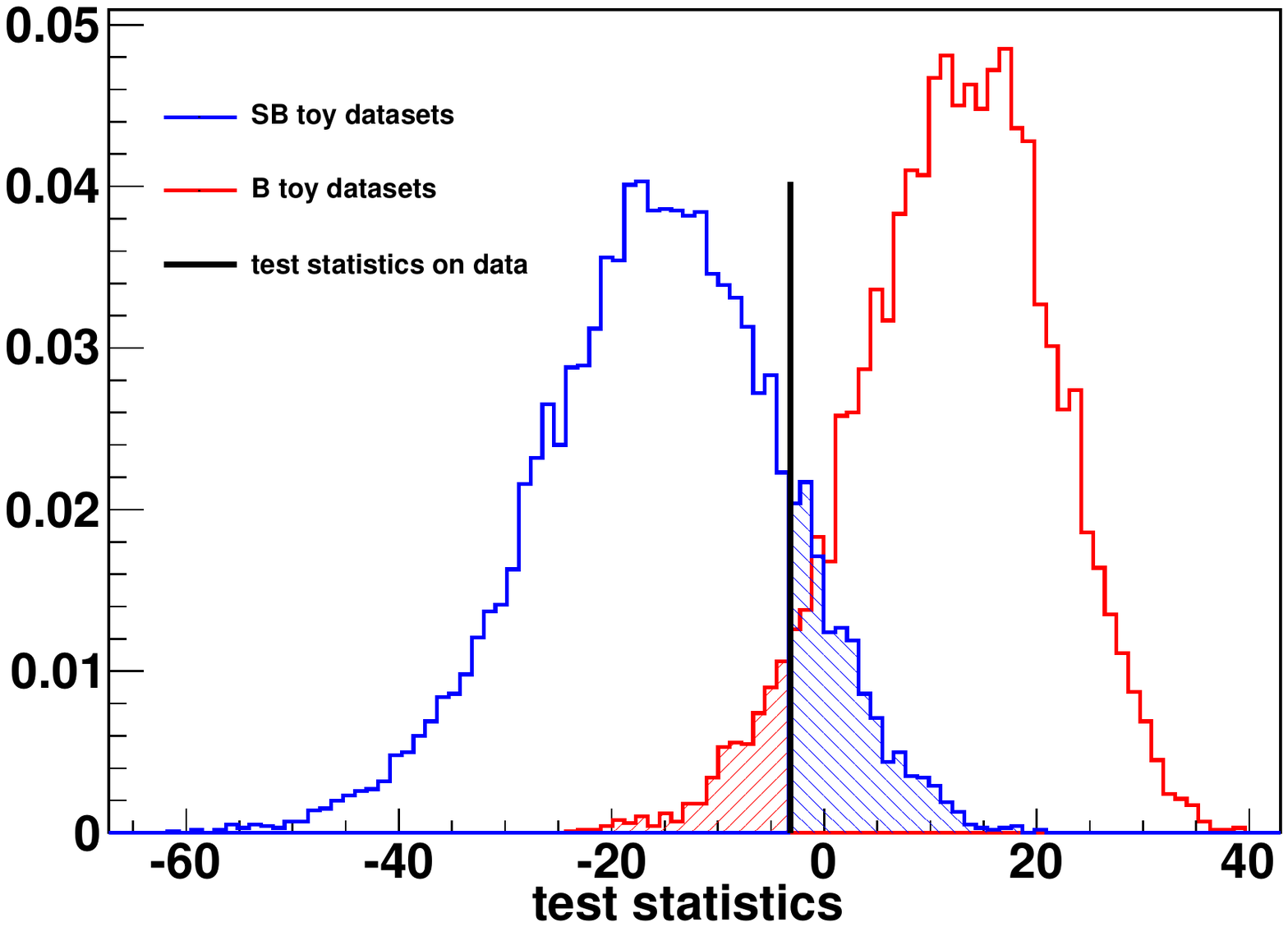} 
        \caption{
        Result from the hybrid calculator, the distributions of $-2\ln Q$ in the background-only
                 (red, on the right) and signal+background (blue,
                 on the left) hypotheses. The black line represents
                 the value $-2\ln Q_{obs}$ on the tested data.  The
                 shaded areas represent $1-CL_{b}$ (red) and $CL_{sb}$
                 (blue).
                 \label{m2lnQ} }
  \end{center}
\end{figure}

From the comparison of the test statistics value of $Q$, observed in the data ($Q_\mathrm{obs}$), to the
probability distribution of $Q$   ($dP/dQ$ )
expected for the $b$ and $s+b$ hypotheses,  the confidence levels, $CL_{sb}$ and $CL_{b}$, can be computed. 
Since the functional forms of the test statistic distributions 
$dP_{sb}/dQ$ and $dP_{b}/dQ$  
are not always known a priori, a large amount of toy Monte Carlo experiments are performed. 
Two types of pseudo datasets are generated: with signal plus background and with 
the background-only hypothesis. Figure~\ref{m2lnQ} provides an example of the obtained distributions from the 
two pseudo data sets and how they compare with the observed test statistic value in the data.

Systematics uncertainties are taken into account  using Bayesian
Monte Carlo sampling.
For each toy Monte Carlo experiment, the effective value 
of the nuisance parameters is varied before generating the toy sample. 
The whole phase space of the nuisance parameters is thus sampled through
Monte Carlo integration. The final net effect consists in a broadening
of the test statistic distribution and thus, as expected in presence of
systematic uncertainties, a degraded separation of the hypotheses.
\par
All this is implemented in Roostats  by the {\tt HybridCalculator} class. 
Input to the class are the model for the null hypothesis ($s+b$) and the model for the alternate ($b$ only) hypothesis, 
the data sets and optionally the prior distribution for the nuisance parameters which is sampled in the toy generation process.  
The {\tt HybridCalculator} class  provides various
choices for the test statistics, such as the number of events,  the likelihood ratio or the profiled likelihood ratios.
The results of the  {\tt HybridCalculator} consists of the test statistic distribution for the two hypothesis, from which the hypothesis p-value and significance can be obtained. 
Different results can be merged together in order to be able to run the calculator in distributed computing processes. The {\tt HybridPlot} class provides a way of plotting the result, as shown for example in figure~\ref{m2lnQ}. 
 
\par
By varying the parameter of interest representing the 
hypothesis being tested (for example by varying the signal cross-section) one can obtain 
a one sided confidence interval (i.e. an exclusion limit). 
RooStats provides a class, {\tt HypoTestInverter}, which  implements the interface {\tt IntervalCalculator} and performs the scanning of the hypothesis test results of the
 hybrid calculator for various values of one parameter of interest.
 By looking where the confidence level curve of the result intersects the desired confidence level,  an upper limit can be derived, assuming the interval is connected.

\section{Combination of Results}

Combining results from multiple experiments in order to enhance sensitivity of a measurement or improve the power of a hypothesis test is common.  The challenge of combining results is primarily logistical, since a proper combination requires low-level information from each experiment be brought together to form one large statistical test. Again, this is hindered by the fact that the ingredients to the combination are heterogeneous (eg. different formats, technologies, and conventions).  
A major advancement that was made by the RooStats project is the concept of the workspace.   The power of the workspace is that it allows one to save data and an arbitrarily complicated model to disk in a ROOT file.  These files can then be shared or archived, and they provide all the low-level ingredients necessary for a proper combination in a unified framework.   A direct advantage of this is a digital publishing of the results. 

The {\tt RooWorkspace} class of RooFit provides the low-level functionality for storing the full model and the data and,  in addition, it provides a convenient functionality to create easily the model via a string interface (workspace factory). 


\section{RooStats Utilities}

In addition to the previously described methods for estimating confidence intervals and performing hypothesis tests, RooStats provides also some statistical  utilities. These include functions to calculate the p-value or Z-value (e.g. significance in one-sided Gaussian standard deviations)  for a number counting experiments. 

Another utility is the class {\tt RooStats::SPlot}, which implements the  $\hbox{$_s$}{\cal P}lot$ technique~\cite{splot} to disentangle signal from background via an extended maximum likelihood fit and with a tool to access the quality and validity of the fit by producing distributions for the control variables.  {\tt RooStats::SPlot} complements the ROOT {\tt TSPlot}\footnote{see http://root.cern.ch/root/html/TSPlot.html} class with the possibility to use arbitrary models created with RooFit.

The {\tt BernsteinCorrection} is another utility class to augment a nominal probability density function with a polynomial correction term. This is useful for incorporating systematic variations to the nominal probability density function.   The Bernstein basis polynomials are particularly appropriate because they are positive definite. 

The {\tt HLFactory} is a user friendly wrapper around the RooFit workspace class to build the input model for RooStats from a text file. Other high level factory tools are planned and will available in future versions of RooStats.  

\section{Conclusions}

RooStats is a new software package providing the major statistical tools required for the LHC data analysis. 
These tools are implemented by using common interfaces, so they can be re-used with the same input model and they can be applied  both in simple number counting problems and in complex ones, with multi parameter of interests and various nuisance parameters. 

RooStats is distributed together with ROOT and its latest production version, released in December 2009, is 5.26.00. Examples of usage are provided  in the {\tt tutorials/roostats}  directory of the ROOT distribution as ROOT macros.   
There are macros as example for each of the RooStats calculator. In addition to numbers (interval limits, significance, etc..) they produce also statistical plots 
of the result (confidence level contours or test statistics distributions like in figure~\ref{m2lnQ}).  The examples are based on models with Poisson statistics (number counting experiments) or Gaussian signal model with flat background and they take into account  systematic uncertainties in the nuisance parameters. Furthermore, a tutorial of the {\tt FeldmanCousins} class is provided, which uses the same model of the neutrino oscillation search described in~\cite{fc_paper}. 

RooStats is starting now to be used by both ATLAS and CMS for producing the results of their data analysis and it is then planned to be used in the future for their analysis combinations.  

\section*{Acknowledgements}
We are thankful to the member of the ATLAS and CMS statistics committees
for the exchange of ideas, advice and encouragement. We also wish to
thank R. Cousins, G. Piacquadio, M. Pierini, G. H. Lewis and M. Pelliccioni  which have contributed to the developments of  RooStats.


\begin{thebibliography}{99}

\bibitem{kyle} K. S. Cranmer, \emph{Statistics for the LHC: Progress, Challenges and Future}, proceedings of the PHYSTAT LHC Workshop, CERN 27-29 June 2007, 47. 

\bibitem{roostats_cms} D. Piparo, G. Schott and G. Quast, \emph{RooStatsCms:: a tool for analysis modelling, combination and statistical studies},  J. Phys.: Conf. Ser. 219 032034 
(2010)

\bibitem{ROOT} I. Antcheva et al, \emph{ROOT - A C++ framework for petabyte data storage, statistical analysis and visualization}, Comput. Phys. Comm. 180 (2009)
2499.

 \bibitem{roofit}W. Verkerke, \emph{Statistical Software for the LHC}, proceedings of the PHYSTAT LHC Workshop, CERN 27-29 June 2007, 169. 

\bibitem{James_book}F. James, \emph{Statistical Methods in Experimental Physics 2$^\mathrm{nd}$ Edition}, Word Scientific 2006.

\bibitem{Minuit} F. James, \emph{MINUIT Reference Manual}, CERN Program Library Writeup D506.

\bibitem{bat}  A.Caldwell, D. Kollar, K. Kr\"{o}ninger, ``BAT - The Bayesian Analysis Toolkit'', arXiv:0808.2552.

\bibitem{fc_paper} G. Feldman and R. D. Cousins, \emph{Unified approach to the classical statistical analysis of small signals}  Phys. Rev. D 57, 3873 (1998)

\bibitem{Cranmer2003}
K.~S.~Cranmer,  \emph{Frequentist hypothesis testing with background uncertainty},
  procedings of PhyStat 2003, SLAC (2003)  [physics/0310108].
  
 \bibitem{cousins-highland} R.D. Cousins and V.L. Highland, ``Incorporating Systematic Uncertainties into an Upper Limit'', Nucl. Instrum. Meth. A320 (1992) 331.

\bibitem{splot} M. Pivk and F. R. Le Diberder, Nucl. Inst. Meth.A 555, 356 (2005)

\end{thebibliography}
\end{document}